\documentclass[preprint,prb,a4paper,showpacs,floatfix]{revtex4}

\usepackage{latexsym}
\usepackage{amsmath}

\usepackage[dvips]{graphicx}
\graphicspath{{./figs/}}

\usepackage{subfigure}
\usepackage{color}
\usepackage[latin1]{inputenc}

\newcommand {\qref}[1]{Ref.~\onlinecite{#1}}
\newcommand {\queq}[1]{(\ref{#1})}
\newcommand {\qeq}[1]{Eq.~\queq{#1}}
\newcommand {\qeqs}[1]{Eqs.~\queq{#1}}

\newcommand {\Eth}{E_{\rm th}}

\newcommand {\eps}{\epsilon}
\newcommand {\epsc}{\epsilon_c}

\newcommand {\Art}{Ar$_{1000}$}
\newcommand {\Cut}{Cu$_{1000}$}

\begin{document}

\date{\today}

\title{Cluster-induced crater formation}

\author{Christian Anders}
\author{Gerolf Ziegenhain}
\author{Steffen Zimmermann}
\author{Herbert M.~Urbassek}
 \email{urbassek@rhrk.uni-kl.de}
 \homepage{ http://www.physik.uni-kl.de/urbassek/}
\affiliation{%
Fachbereich Physik, Universit{\"a}t Kaiserslautern, \\
Erwin-Schr{\"o}dinger-Stra{\ss}e, D-67663 Kaiserslautern, Germany}

\begin{abstract}

Using molecular-dynamics simulation, we study the crater volumes induced
by energetic impacts ($v= 1- 250$ km/s)
of projectiles containing up
to $N=1000$ atoms. We find that for Lennard-Jones bonded material the
crater volume depends solely on the total impact energy $E$. Above a
threshold $\Eth$, the volume rises linearly with $E$. Similar results
are obtained for metallic materials. By scaling the impact energy $E$ to
the target cohesive energy $U$, the crater volumes become independent of
the target material. To a first approximation, the crater volume
increases in proportion with the available scaled energy, $V=aE/U$. The
proportionality factor $a$ is termed the cratering efficiency and
assumes values of around 0.5.

\end{abstract}


\maketitle


\section{Introduction}

The erosion of surfaces by atom or ion impact -- i.e., the sputter
process -- has for long been studied.\cite{BehI,BehII,BehIII,BE07} More
recently, interest has focussed on erosion by cluster impact both
experimentally\cite{ABD*98,BDD*01,BBD*02,AJOT03,BD04} and by computer
simulations.\cite{IY99,ST04,CU00,CAKU00,YG02,SNK03,ZU05} We shall
investigate in this paper the question how the crater volume depends on
the cluster energy and cluster size $N$ and, -- more specifically --
whether it is the total energy $E$ of the cluster, or rather the energy
per atom $E/N$, which is decisive to determine the cluster volume. We
shall employ two widely differing classes of materials to study this
question, a van-der-Waals bonded target, and metals, in order to show in
how far our considerations are materials independent.

\section{Method}

We employ the method of molecular-dynamics simulation to shed light on
the process of crater formation. This technique is standard, and will
not be presented here. Details are given
elsewhere.\cite{CAKU00,AU00,SUZ02,AUJ04,UU06} In short: The clusters are
chosen of a spherical shape and consist of $1 \le N \le 1000$ atoms. For
the Ar system, a Lennard-Jones potential,\cite{MWW49,HV69} and for the
Cu and Au targets a many-body potential of the embedded-atom
type\cite{DFB93} has been chosen.\cite{MFMP99,CU00,ZU05} In all cases,
the potentials have been splined to an appropriate high-energy
potential\cite{WHB77,ZBL85book} in order to accurately model close
collisions. The size of the target system varies between $7 \times 10^4$
and approximately $7 \times 10^6$ atoms, depending on the total cluster
energy $E$. At the lateral and bottom sides of the simulation target, we
employ damped boundary conditions in order to mimic energy dissipation
to the surrounding target material. The Cu target consists of an fcc
crystal with (100) surface; for the Au crystal a (111) surface has been
chosen. In the case of Ar, we employ an amorphous
target. We determine temperature and pressure in our simulation as local
quantities, which are averaged over a sphere with a radius equal to the
cutoff radius of the potential (containing around 50 atoms) to reduce
fluctuations.\cite{CU97}

\section{Results}

Fig.~1 displays the results of \Art\ impact on an amorphous Ar surface
at 4 keV impact energy. A compression wave moves hemispherically out of
the impact point. The material within the immediate impact zone is seen
to have gasified; this process still continues at the time of $t=3$ ps,
where the snapshots are displayed. Note that the temperatures in the
central region are high, far above the melting and even the boiling
point of this material. The latest snapshot shown ($t=60$ ps)
demonstrates that the crater has considerably widened. The relatively
high temperatures present indicate that the crater form will still relax
to some degree after this time. The simulation results show that the
sputtering process corresponds to a {\em phase explosion}, in which
sputtering occurs by the gasification of the high-energy-density zone,
as long as this is situated sufficiently close to the surface.

The impact of a \Cut\ cluster on a Cu target is displayed in Fig.~2.
Here the crater formation is a faster process, and hence we display
atomistic snapshots already at time of $t=1$ ps. Temperatures do not
reach so high values, when compared to typical materials parameters
such as the triple or critical temperature; the crater walls are molten,
but the boiling point
is not reached. However, the pressure reaches high values: Immediately
below the crater, a zone of high compressive pressure has formed; its
anisotropy reflects the crystallinity of the target. Close to the
surface, we observe already a zone of tensile pressure forming; the
further evolution of this zone will be discussed elsewhere. At the final
time displayed, the form of the crater seems to have stabilized; the
temperature is close to zero. Note that the crater has apparently shrunk
after $t=1$ ps in the course of the target relaxation.

We define the crater volume as the ensemble of missing atoms below the
original surface. Consequently, we measure the crater volume $V$ as a
dimensionless quantity, viz., the equivalent number of missing target
atoms. The total kinetic energy of the impacting cluster, $E$, will be
scaled to the target cohesive energy, $U$, and is thus measured as a
dimensionless energy

\begin{equation}  \label{eps}
\epsilon =  E / U .
\end{equation}

For the materials used in our study, it is $U=0.0815$ (3.54, 3.79) eV
for Ar (Cu, Au). Fig.~3 summarizes the energy dependence of the crater
volumes
induced in the two materials studied. Analogous results for smaller
cluster size
$N=100$ have been published previously.\cite{UASU08}
In both cases,
self-bombardment by clusters containing 1000 atoms has been simulated.
Evidently the crater volumes for these two widely different materials
coincide rather well when the impact energy is scaled to the target
cohesive energy, $\eps=E/U$. The data are -- to a good first
approximation -- well described by a linear law

\begin{equation}  \label{V}
V = a (\eps-\eps_c), \quad \eps > \eps_c ,
\end{equation}

where the \emph{cratering efficiency} $a \cong 0.5 $, Ar (Cu), and the
threshold energy is $\eps_c \cong 4700$. More precisely, a linear fit to
our data gives $a = 0.52 \pm 0.02$ and $\eps_c=4725 \pm 480$ for Ar,
while the fit for Cu yields $a =0.47 \pm 0.04$ and $\eps_c=4725 \pm
1570$.

We rationalize the simple law, \qeq{V}, in which only one materials
parameter, the cohesive energy $U$, describes the physics, as follows:
The cluster is quickly stopped in the target, on a time scale $t_0 \cong
d/v$, where $d$ is the cluster diameter, and $v$ its impact
velocity.\cite{AU07} After this time, virtually all the cluster energy
$E$ is available close to the target surface for crater formation. The
available energy can then be used for bond breaking in the target and
hence atomize the material in the energized region, which is to become
the crater volume.

Finally, Fig.~4 assembles the simulated crater sizes from the
present simulations and combined with a larger set of previous
simulations on small Cu clusters ($N=13$, 43).\cite{AU00} Data for Au
cluster impacts are also shown, which have been extracted from our
previous simulations.\cite{ZU05,ZU07} These latter data
are fitted to a law

\begin{equation}  \label{Joh}
V = c \frac{\eps^{1+d}}{(\eps + \epsc)^d} ,
\end{equation}

with $c=0.511$, $\epsc = 1320$, and $d=1.5$. Such a law may be better
suited to describe
the threshold behaviour, while for large $\eps$, it again leads to a
linear increase.\cite{AUJ04}  Fig.~4 demonstrates that in the energy
regime studied here, the linear regime describes well crater volumes
both in condensed noble gases and metal target. The threshold regime,
however, is dependent on materials and, in particular, on the cluster
size.

In Fig.~5, we plot the same data as a function of the energy per
particle, $E/N$. Note that for a fully linear volume-energy
relationship, with size independent parameters, again all data should
converge to one universal line. We see that for high impact velocities,
$\eps/N \agt 100$, this is indeed the case. In the threshold regime,
however, the data show an increasingly strong dependence on cluster
size.

\section{Conclusion}

\begin{enumerate}

\item Molecular-dynamics simulations of cluster-induced crater volumes
$V$ give comparable results for different target materials if the
cluster energy $E$ is scaled to the target cohesive energy $U$.

\item Above a threshold $\Eth$, the crater volume $V$ increases linearly
with the cluster energy $E$.

\item Crater formation sets in when the excitation strength exceeds a
certain threshold. This threshold is mainly characterized by an energy
criterion, such that the cluster impact energy scaled to the cohesive
energy of the target must exceed a threshold value, which is only mildly
dependent on the material. These thresholds attain similar values, even
for so drastically different materials as van-der-Waals bonded materials
and metals.

\end{enumerate}

\bibliography{submitted}

\begin{figure}
\begin{center}
\subfigure[]{\includegraphics*[width=0.3\linewidth]{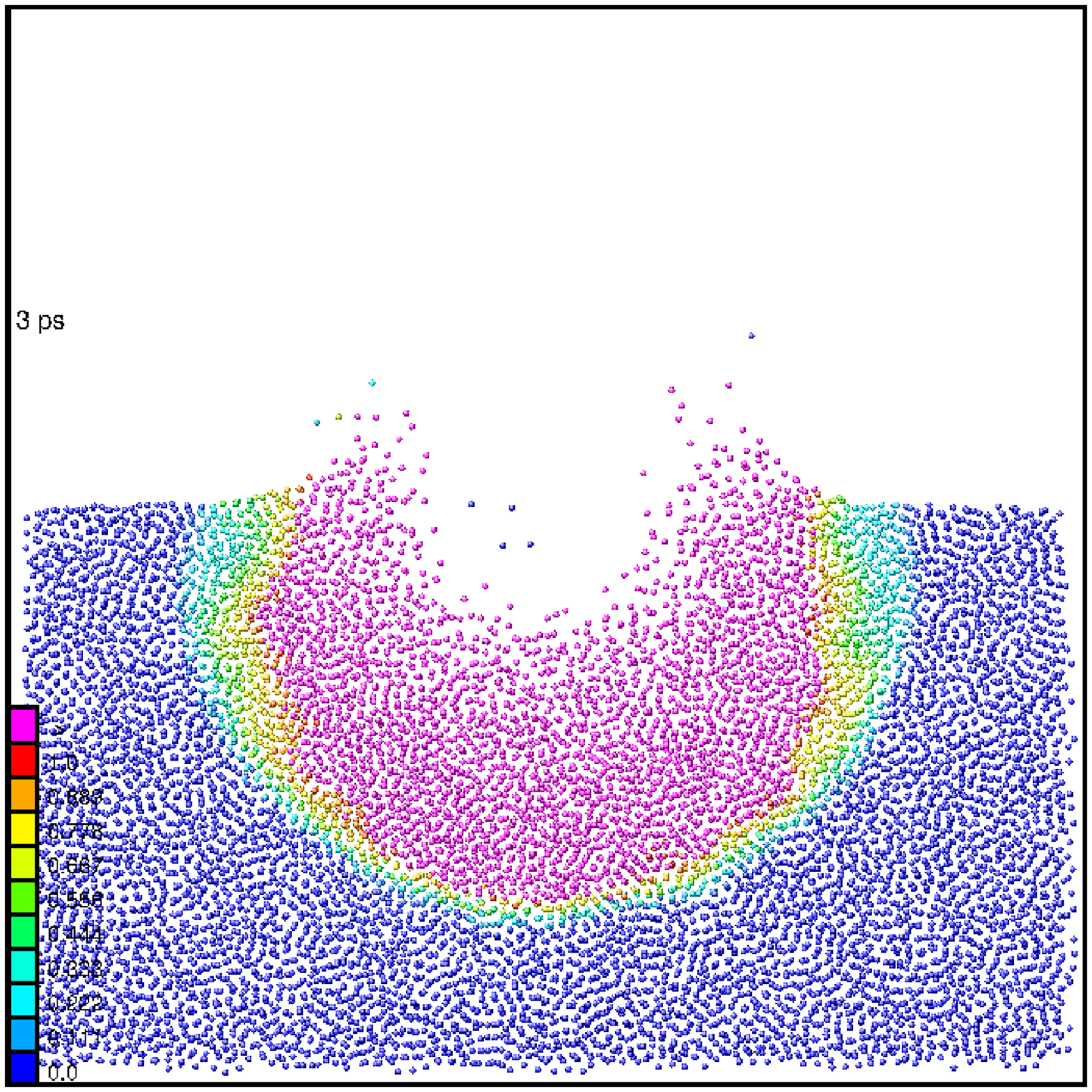}} \hfill
\subfigure[]{\includegraphics*[width=0.3\linewidth]{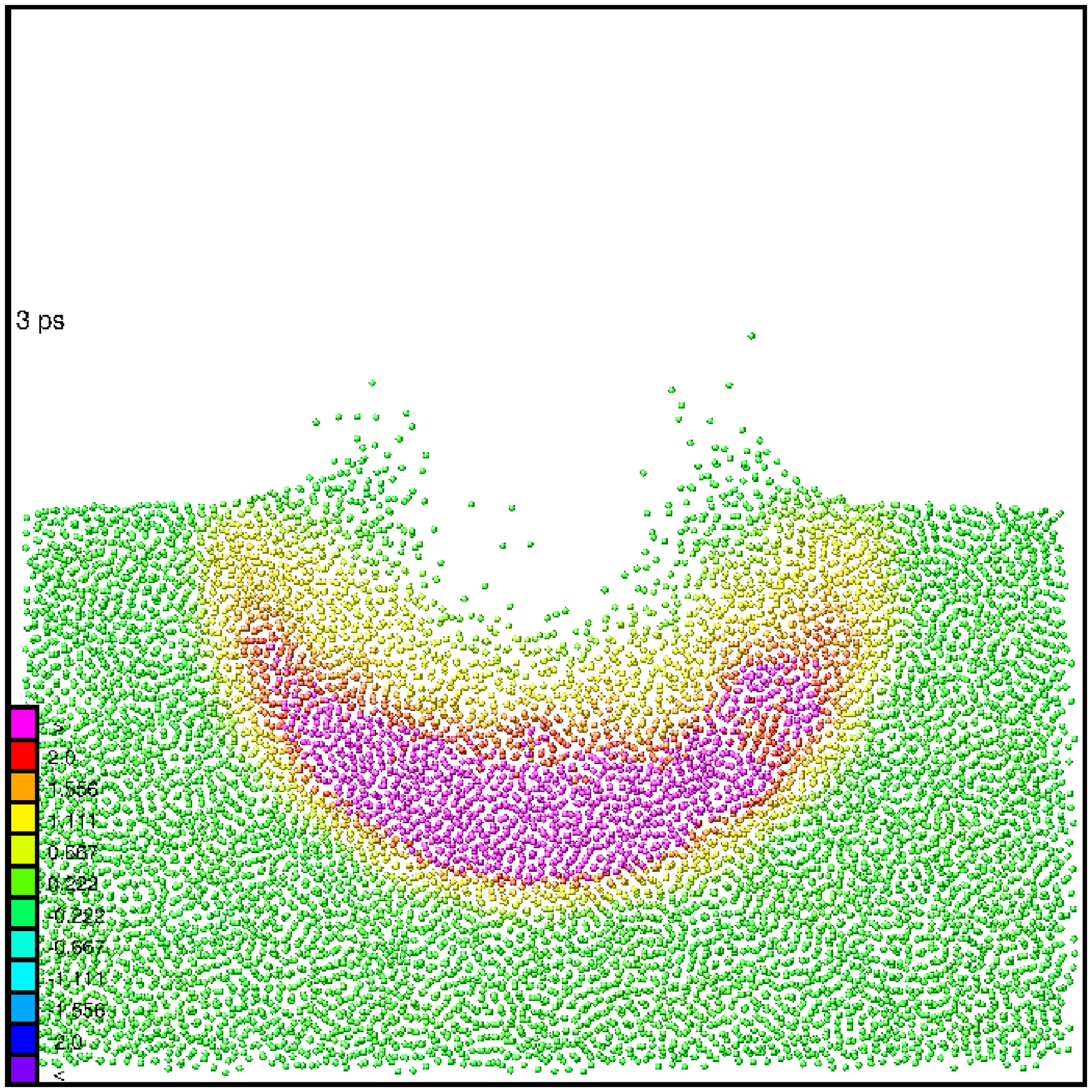}} \hfill
\subfigure[]{\includegraphics*[width=0.3\linewidth]{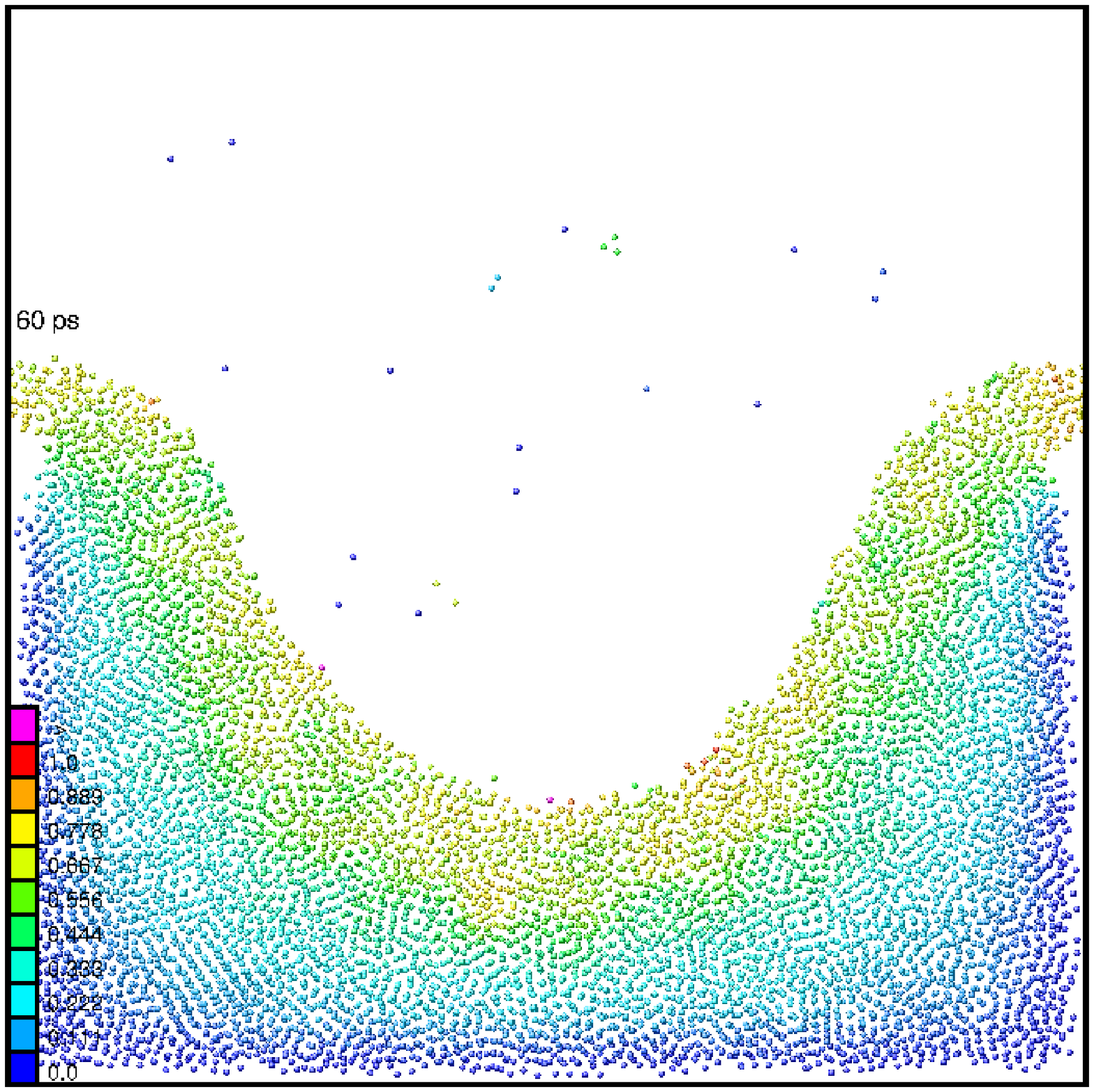}}
\end{center}

\caption{ Snapshots of \Art\ $\to$ Ar impact at an impact energy of
$E=4$ keV, $\eps \cong 49,000$.
a: Temperature distribution in the target at $t=3$ ps after cluster
impact. Color denotes temperature in units of the boiling temperature of
Ar (87.3 K). b: Pressure distribution in the target at $t=3$ ps after
impact. Color: pressure in units of GPa.
Green denotes zero pressure, while the highest pressure (purple)
is compressive at $>2$ GPa.  c: Crater formed at $t=60$ ps after impact.
Color denotes temperature as in subfigure (a).
 }

\label{fc_Ar}
\end{figure}

\begin{figure}
\begin{center}
\subfigure[]{\includegraphics*[width=0.3\linewidth]{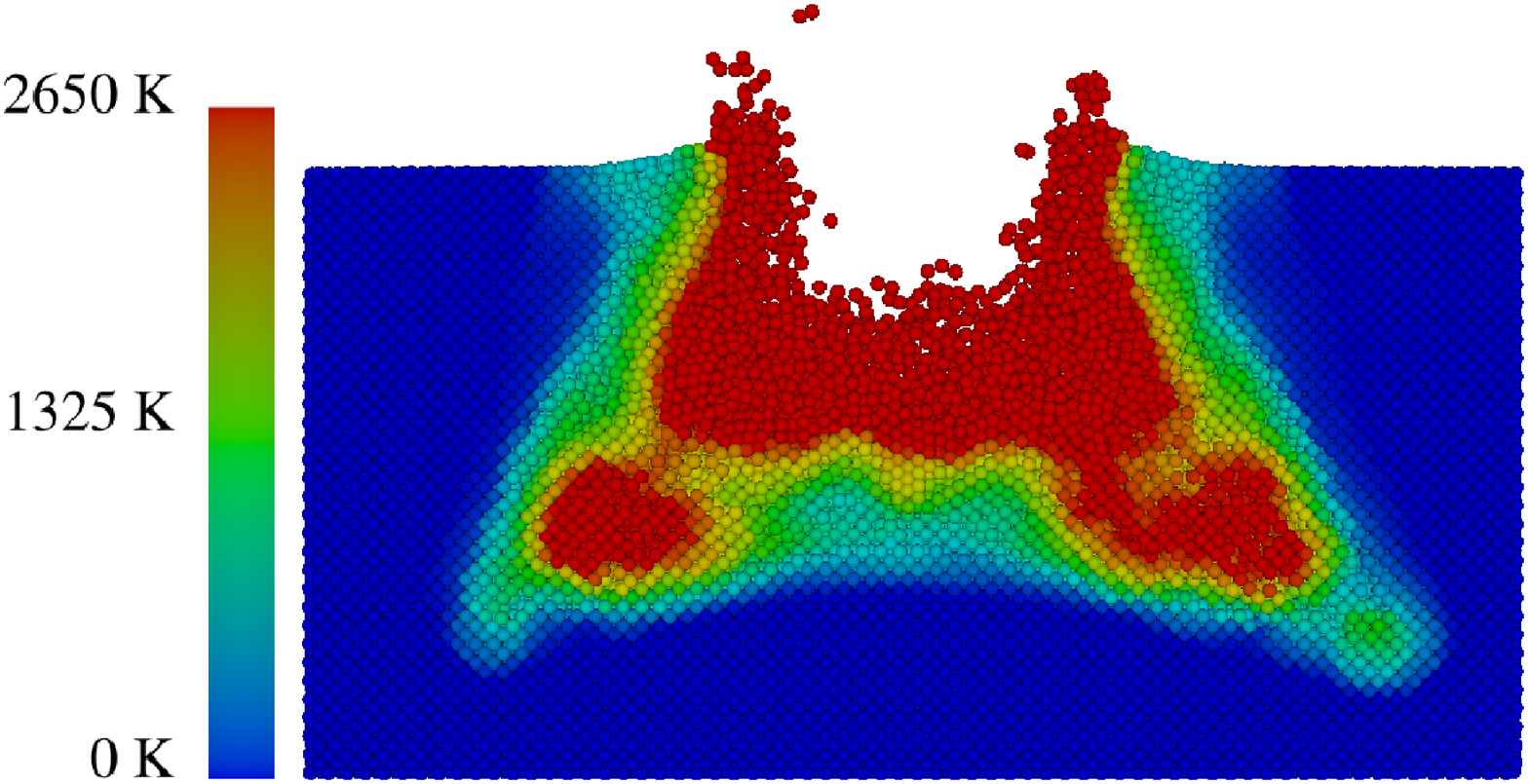}} \hfill
\subfigure[]{\includegraphics*[width=0.3\linewidth]{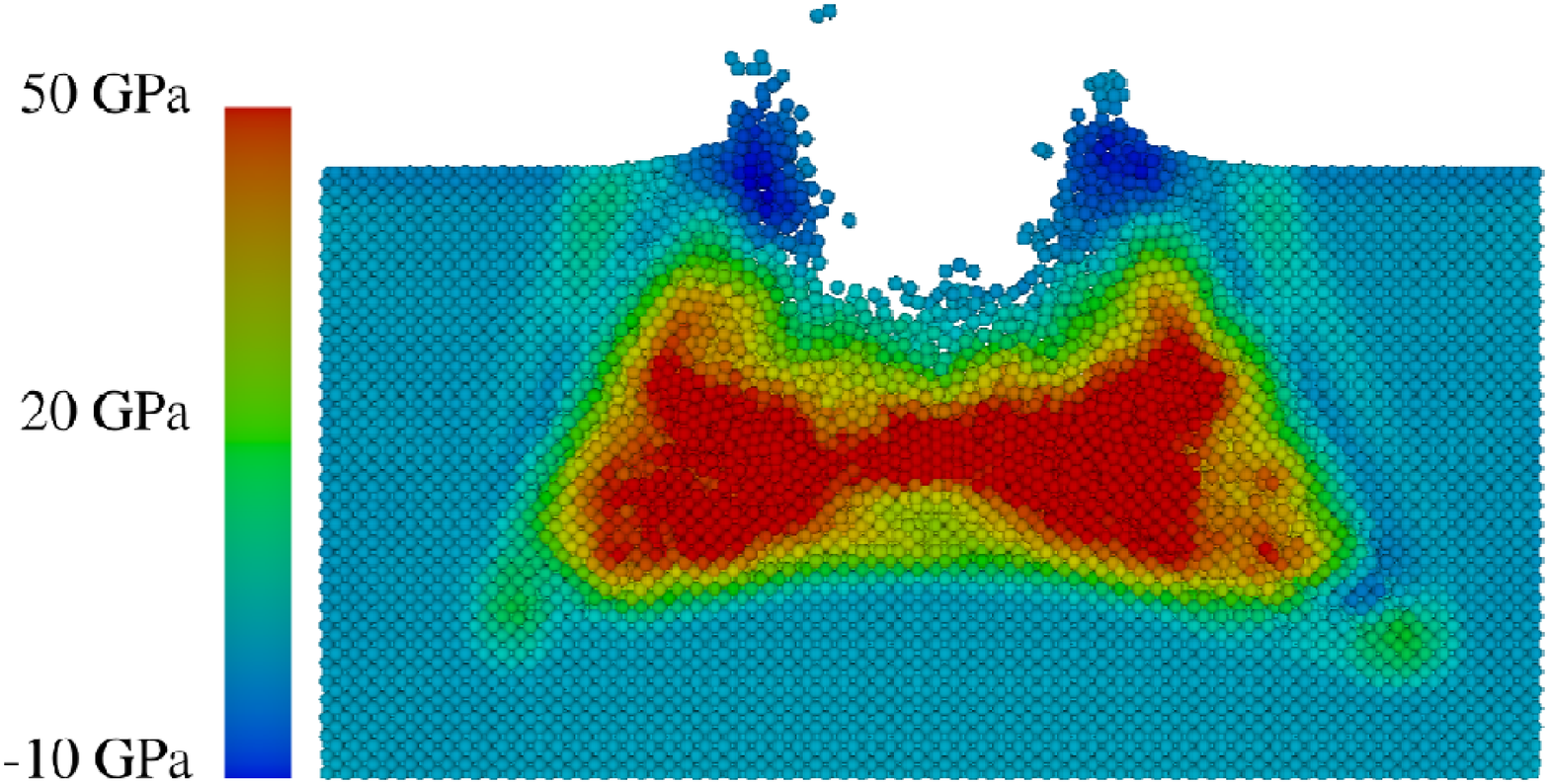}} \hfill
\subfigure[]{\includegraphics*[width=0.3\linewidth]{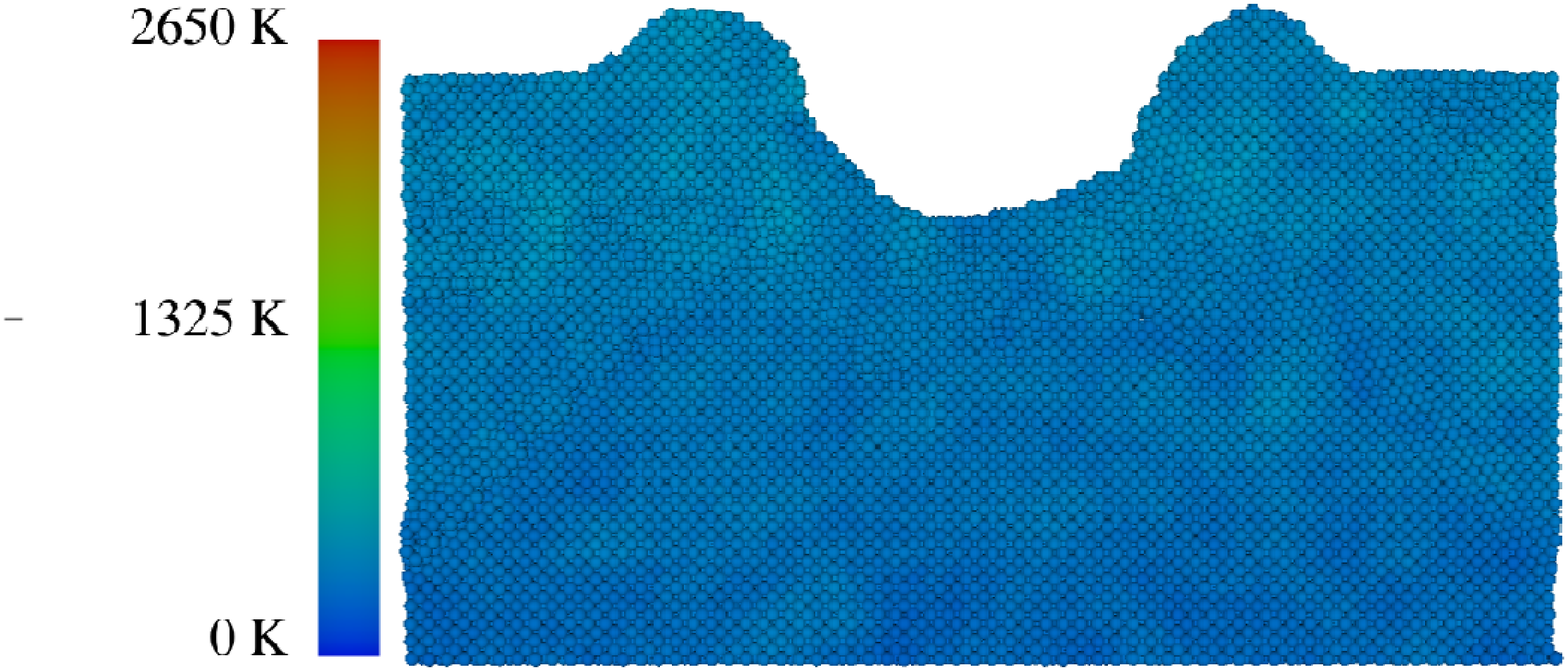}}
\end{center}

\caption{ Molecular-dynamics view of crater formation in a Cu target
bombarded by a \Cut\ projectile at $E=50,000$ eV, $\eps \cong 14,100$.
a:
Temperature distribution in the target at $t=1$ ps after cluster impact.
b: Pressure distribution in the target at $t=1$ ps after
impact.  Turquoise denotes zero pressure,
positive pressure is compressive, while negative pressure is tensile. c:
Crater formed at $t=50$ ps after impact.
 }

\label{fc_Cu} \end{figure}

\begin{figure}
\begin{center}
\includegraphics*[angle=270,width=0.45\linewidth]{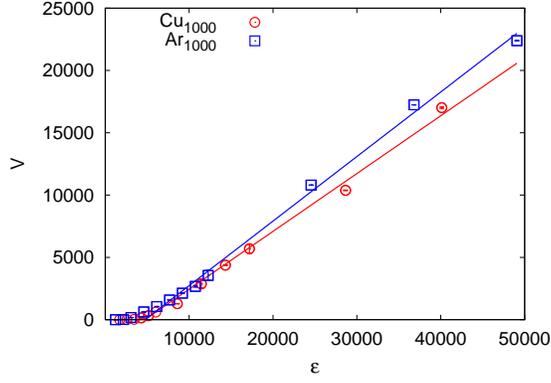}
\end{center}

\caption{  Dependence of the cluster volume $V$ on the scaled impact
energy, $\epsilon =  E / U$. Self-bombardment of clusters containing
$N=1000$ atoms on Ar and Cu has been simulated.
 }

\label{f4}
\end{figure}

\begin{figure}
\begin{center}
\subfigure[]{\includegraphics*[angle=-90,width=0.45\linewidth]{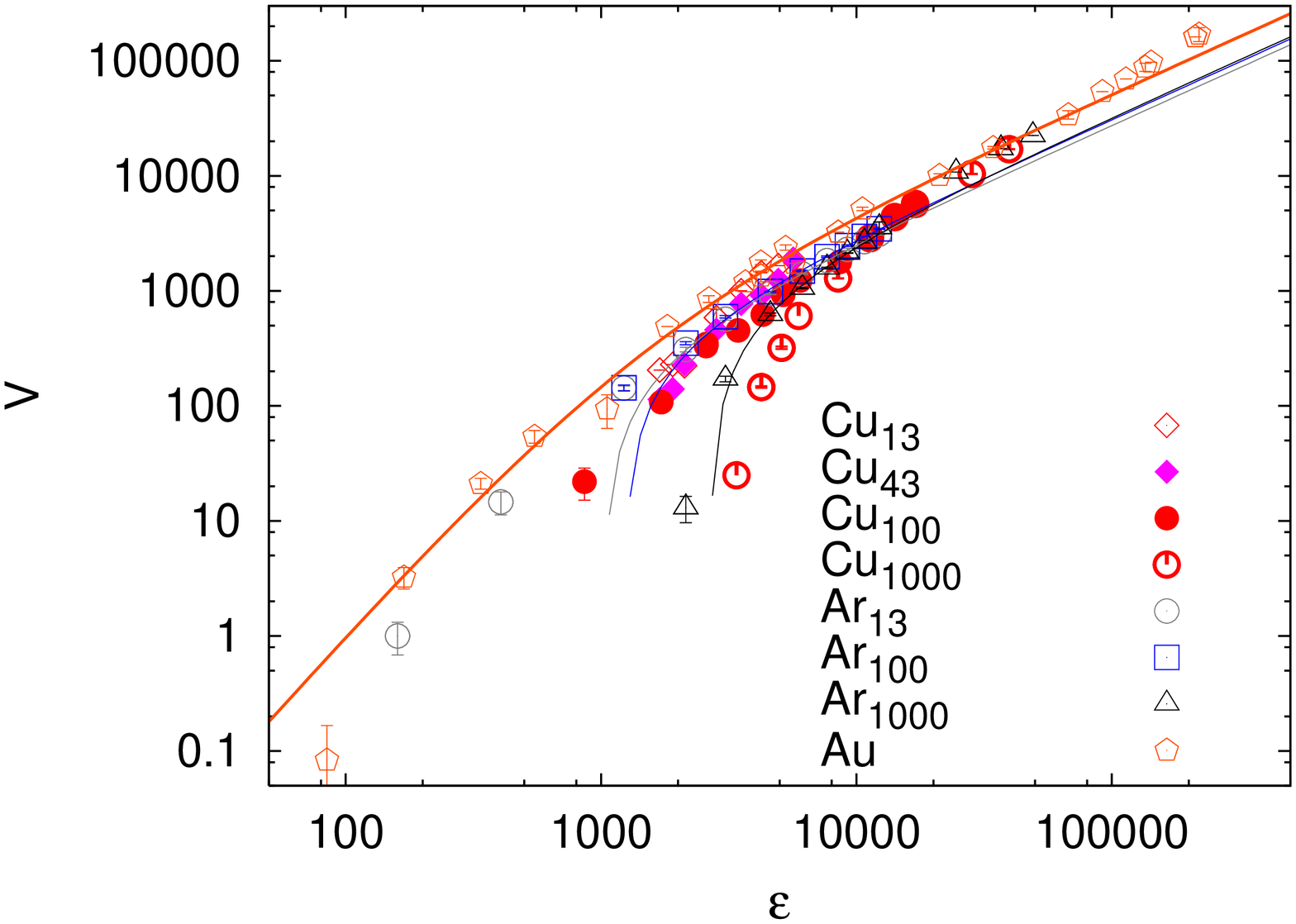}} \hfill
\subfigure[]{\includegraphics*[angle=-90,width=0.45\linewidth]{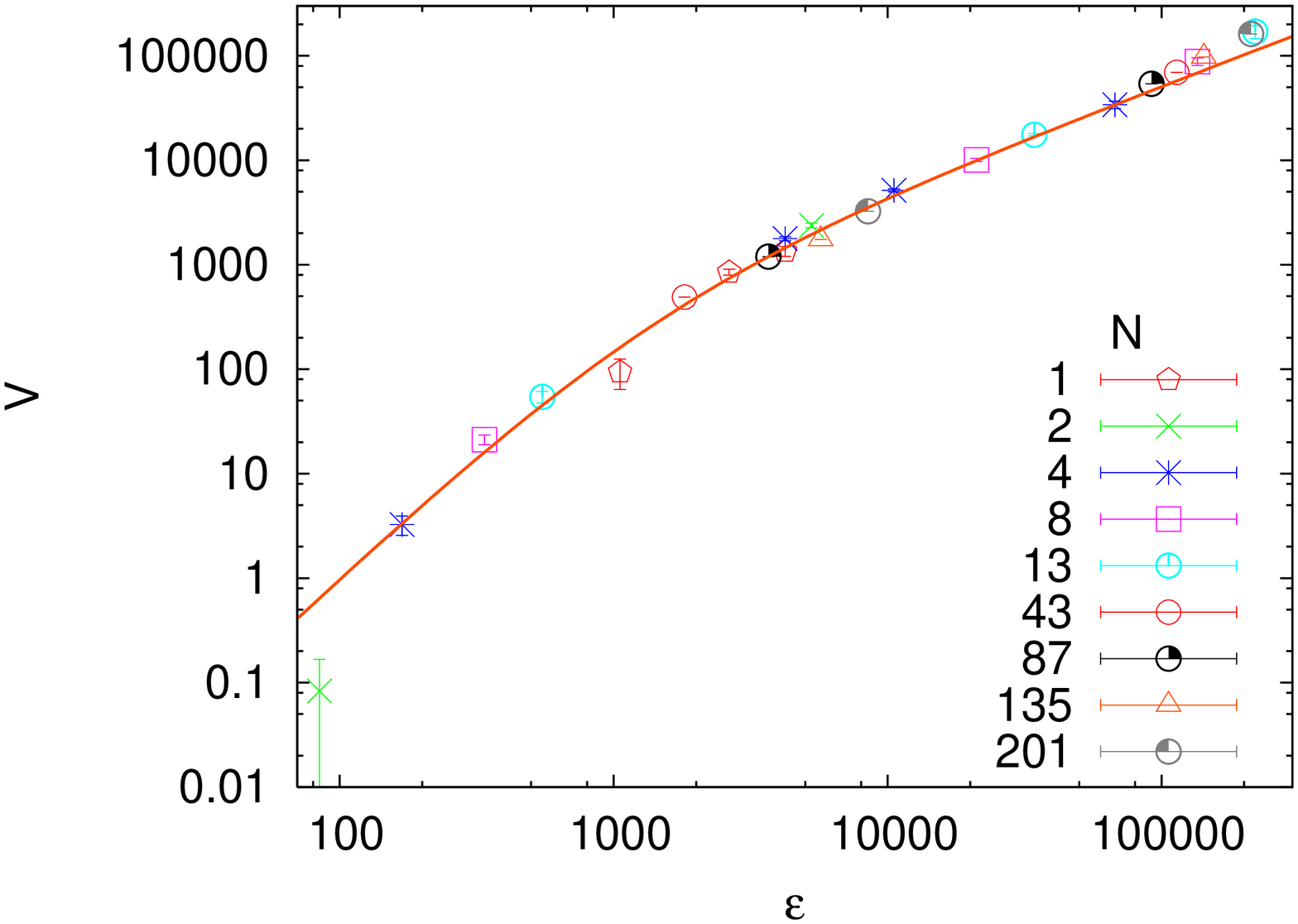}}
\end{center}

\caption{ Synopsis of simulational data of crater volumes $V$ vs scaled
energies $\eps=E/U$. Data  for small Cu clusters ($N=13$,
43) taken from \qref{AU00}.
Legend indicates projectiles. Lines indicate (asymptotically) linear
relationships, \qeqs{V} and \queq{Joh}. Lines in subfigures (a) and
(b) are identical.  (b) details the energy and size dependence of the Au
data of subfigure (a).
 }

\label{f5}
\end{figure}

\begin{figure}
\begin{center}
\subfigure[]{\includegraphics*[angle=-90,width=0.45\linewidth]{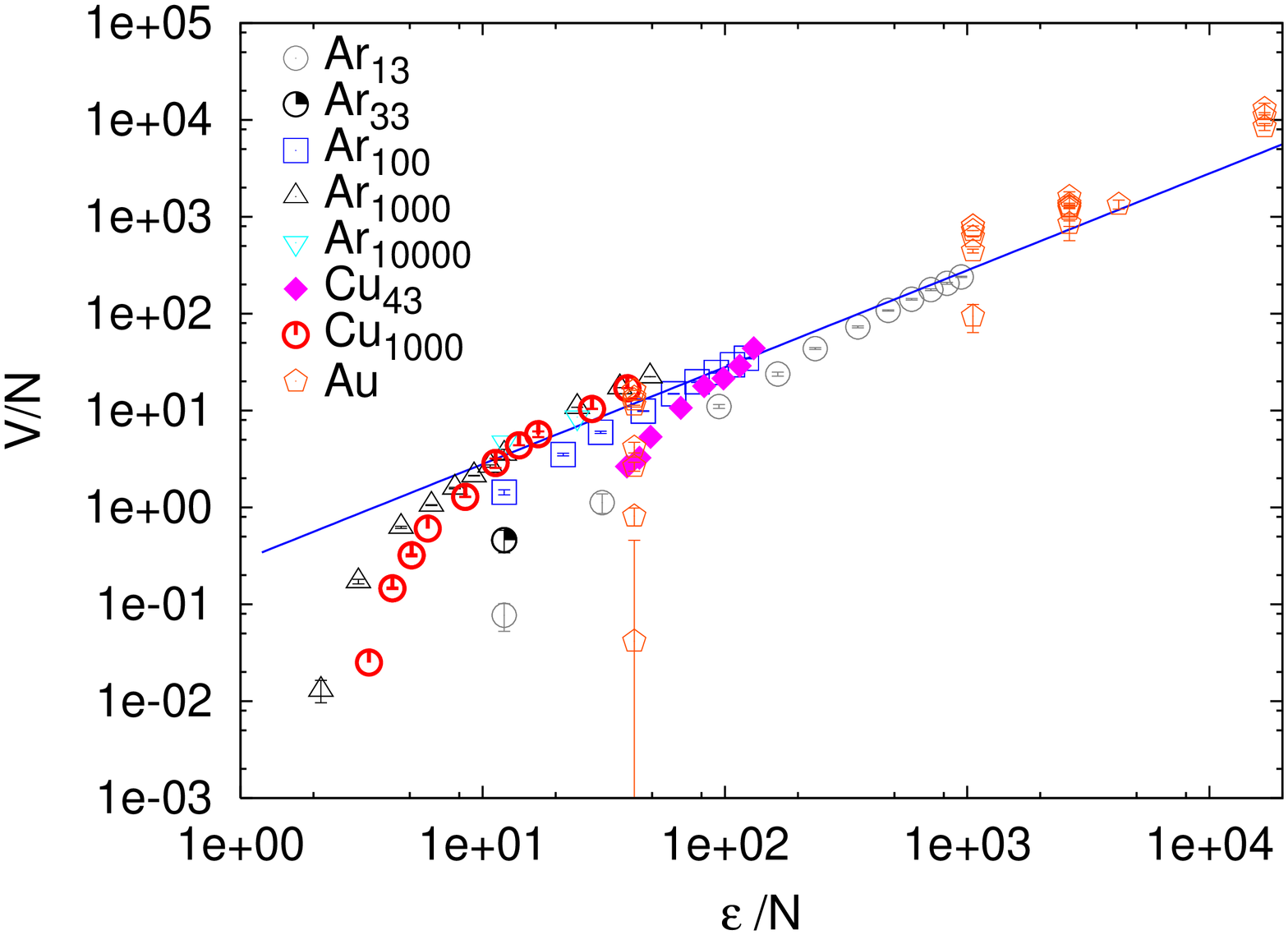}} \hfill
\subfigure[]{\includegraphics*[angle=-90,width=0.45\linewidth]{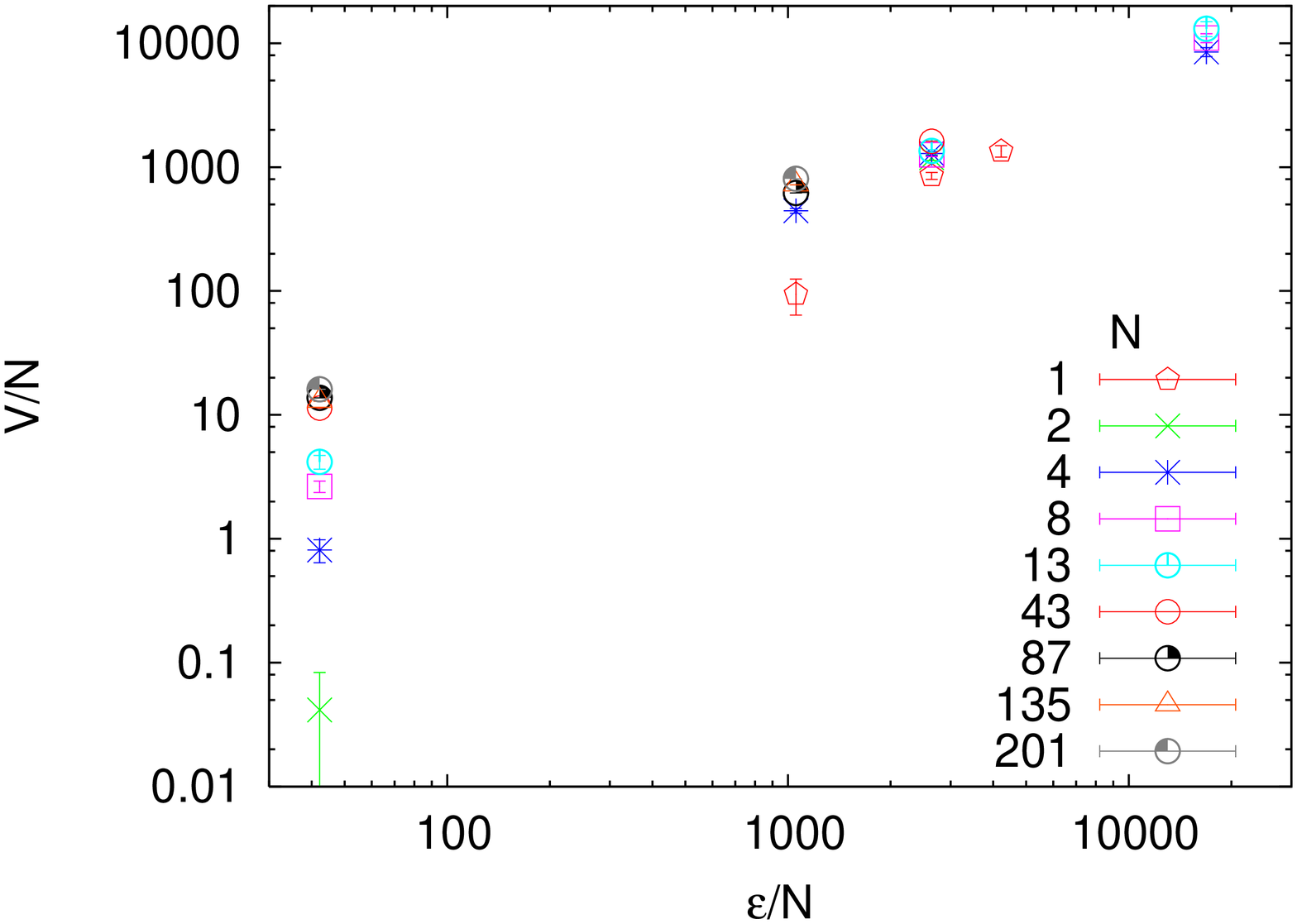}}
\end{center}

\caption{ Data of Fig.~4  scaled to cluster size $N$.
Legend indicates projectiles. Lines indicate a linear relationship.
Lines identical to those in Fig.~4.
 }

\label{f5b}
\end{figure}

\end{document}